# Phonon-mediated superconductivity in silicene

Wenhui Wan, Yanfeng Ge, Fan Yang and Yugui Yao[(a)]

*School of Physics, Beijing Institute of Technology - Beijing 100081, China*

**Abstract** – We predict that electron-doped silicene is a good two-dimensional electron-phonon superconductor under biaxial tensile strain by first-principles calculations within rigid band approximation. Superconductivity transition temperature of electron-doped silicene can be increased to be above 10 K by 5% tensile strain. Band structures, phonon dispersive relations, and Eliashberg functions are calculated for detailed analysis. The strong interaction between acoustic phonon modes normal to the silicene plane and the increasing electronic states around the Fermi level induced by tensile strain is mainly responsible for the enhanced critical temperature.

Silicene, a low-buckled honeycomb sheet of silicon atoms [1-3], has attracted much attention because of the existence of Dirac fermions [4-6], compatibility with the conventional silicon microelectronic technology and edge's resistance to oxygen reactivity [7]. Silicene has also been proved to have nontrivial topological properties such as quantum spin/anomalous Hall effect due to strong spin-orbit coupling [8-10]. Topologically nontrivial state combined with *s*-wave superconductivity (SC) will induce topological SC with Majorana fermion edge modes [11] which can be used as a building block for a topological quantum computer [12]. Hence, silicene may be a promising candidate material for realization of quantum computation provided that the superconductivity and topologically nontrivial state could coexist in silicene.

From the viewpoint of electron-phonon (e-ph) coupling, silicene seems unappealing to form SC considering the heavy atomic mass of silicon. However, boron-doped semiconductors based on group IV elements including diamond, silicon and silicon carbide have been found to be superconductors [13, 14]. In those covalent materials, charges are doped into $sp^3$-bonding bands near the Fermi level that have a large e-ph coupling with the bond stretching modes [15]. It was also predicted that hole-doped graphane had enhanced e-ph coupling because its distortion of carbon sheet favored the $sp^3$ bonding, as diamond, rather than fully $sp^2$ bonding as pristine graphene [16]. The e-ph coupling in silicene may be stronger than graphene because of its intrinsic buckled structure which will lead to partial $sp^3$ hybridization. Furthermore, a simple hexagonal phase of bulk silicon formed under high pressure (18 GPa) was found to be superconducting with the critical temperature $T_c$ = 8.2 K [17, 18]. In addition, $CaSi_2$, a kind of silicide insisting of silicene-like sheets separated by Ca layers, was observed to be superconducting under high pressure and its $T_c$ increased from 4 K to 14 K when the buckling of Si sheets was flattened under increasing pressure ranging from 11 to 20 GPa [19, 20]. So it is expected that silicene may become good conventional superconductor after doping although electronic density of states (DOS) is zero at Dirac point (DP). Excitingly, a possible superconducting gap has recently been observed by Chen *et al.* in silicene epitaxially grown on Ag(111) surface [21]. In Chen's work, Ag substrate plays a role of electron reservoir and shifts the Fermi level of silicene to be 0.5 eV above the DP. In addition, the lattice mismatch between silicene and Ag substrate leads to an obvious strain existing in silicene [21]. Presently there lacks microscopic theory for the SC pairing mechanism in this system, we try to provide an understanding toward it from the perspective of e-ph coupling.

In this work, we report a first-principles calculation on the e-ph interaction of the silicene under biaxial strain within the rigid band approximation [22]. We have calculated the e-ph coupling constant (EPC) $\lambda$ and the superconductivity transition temperature $T_c$ of silicene at various doping levels under different strains. The $T_c$ of electron-doped silicene could reach up to be above 10 K by 5% tensile strain. Enhanced density of states at Fermi level due to tension and strong coupling between out-of-plane acoustic phonon modes and electronic states are the main reasons for the good SC of electron-doped silicene.

We performed density functional theory (DFT) calculations within the local-density approximation (LDA) [23] using the package QUANTUM ESPRESSO [24]. Norm-conserving pseudopotentials [25] were used to describe core-valence interaction. A plane wave basis set with an energy cutoff of 30 Ry and a 30×30×1 Monkhorst-Pack sampling of the Brillouin-zone (BZ) [26] were used for structural relaxations. The convergence criteria for energy and force were set at $10^{-4}$ eV and $10^{-3}$ eV/Å, respectively. The phonon dynamical matrices were obtained on a 15×15×1 grid using density functional perturbation theory (DFPT) [27]. The EPC $\lambda_{q,\nu}$ for phonon with wavevector **q** and branch index $\nu$ was calculated as follows [28]:

[(a)]E-mail: ygyao@bit.edu.cn

$$\lambda_{q,v} = \frac{2}{N_k N(\epsilon_F)\omega_{q,v}} \sum_{k,n,m} |g^{qv}_{k+qm,kn}|^2 \qquad (1)$$
$$\times \delta(\epsilon_{k+q,m} - \epsilon_F)\delta(\epsilon_{k,n} - \epsilon_F),$$

and total EPC $\lambda$ was calculated by $\lambda = \sum_{q,v} \lambda_{q,v} / N_q$. Here $\omega_{q,v}$ is phonon frequency. $\epsilon_{k,n}$ is the energy of electronic state with wavevector **k** and band index $n$. $g^{qv}_{k+qm,kn}$ represents e-ph matrix element. $N(\epsilon_F)$ is the DOS per spin at the Fermi energy $\epsilon_F$. $N_k$ and $N_q$ are the total number of **k** points and **q** points, respectively. The integration over the BZ in eq. (1) was performed with a fine 120×120×1 k-grid. Dirac delta function was replaced by Gaussian broadening function with a smearing width of 0.01 Ry, and the superconducting transition temperature $T_c$ was estimated using McMillan's formula [28]:

$$T_c = \frac{\langle\omega\rangle_{\log}}{1.2}\exp\left(-\frac{1.04(1+\lambda)}{\lambda - \mu^*(1+0.62\lambda)}\right), \qquad (2)$$

where $\langle\omega\rangle_{\log}$ is the logarithmic averaged phonon frequency and the effective Coulomb repulsion $\mu^* = 0.1$ was adopted. Other Coulomb parameters do not affect our conclusions.

The obtained lattice constant $a_0$ of strain-free silicene is 3.83 Å, in agreement with previous results [4, 8, 29]. It has been experimentally reported that biaxial strain can be applied to monolayer graphene by adhering the graphene to the substrate of a small aspect ratio depression [30] or by utilizing piezoelectric actuators [31]. Perhaps similar methods could also be applied to silicene to produce biaxial strain in realistic experiments. Here, biaxial strain is simulated by changing the lattice constant of silicene $a$ and relaxing atomic positions fully. We define the strain as $\varepsilon = |a-a_0|/a_0 \times 100\%$. The lattice stability is examined for both cases of tensile strain and compressive strain through phonon calculations. The lattice instability for silicene is observed for the compression of only 1% because imaginary frequencies appear around $\Gamma$ point of BZ. Similar phenomenon has been reported that silicene would undergo a structural phase transition to form a rhombic $\sqrt{3}\times\sqrt{3}$ buckled superstructure under the compression of about 4% [5]. Thereby we will only consider the situation of tensile strain hereafter. In the strain-free silicene, the Si-Si bond length $d_{Si-Si}$ is 2.25 Å and vertical buckling height $\Delta$ is 0.42 Å. For silicene under 5% tension, $d_{Si-Si}$ increases to 2.33 Å, while $\Delta$ decreases to 0.24 Å.

Band structure and projected density of states (PDOS) of silicene under different tensions are displayed in fig. 1. Fermi level of intrinsic silicene is still at DP located at $K$ and $K^*$ point when the tensile strain is not larger than 5%. Figures 1(a) - 1(c) show a remarkable downshift of the conduction band minimum at the $\Gamma$ point. The PDOS of unstained silicene in fig. 1(d) shows that the DOS among the energy range of ±1.2 eV around DP is mainly from $p_z$ orbitals of Si atom. However, the contributions from $s$, $p_x$ and $p_y$ orbitals to the total DOS above DP are clearly enhanced by the tensile strain as shown in figs. 1(e) - 1(f).

The location of Van Hove singularity is labeled both in band structure and PDOS. When $\epsilon_F$ is at this singularity, the corresponding electron concentration $n_e$ is about $1.8\times10^{14}$ cm$^{-2}$, $2.0\times10^{14}$ cm$^{-2}$ and $3.1\times10^{14}$ cm$^{-2}$ for the tension of $\varepsilon = 0\%$, 3% and 5%, respectively.

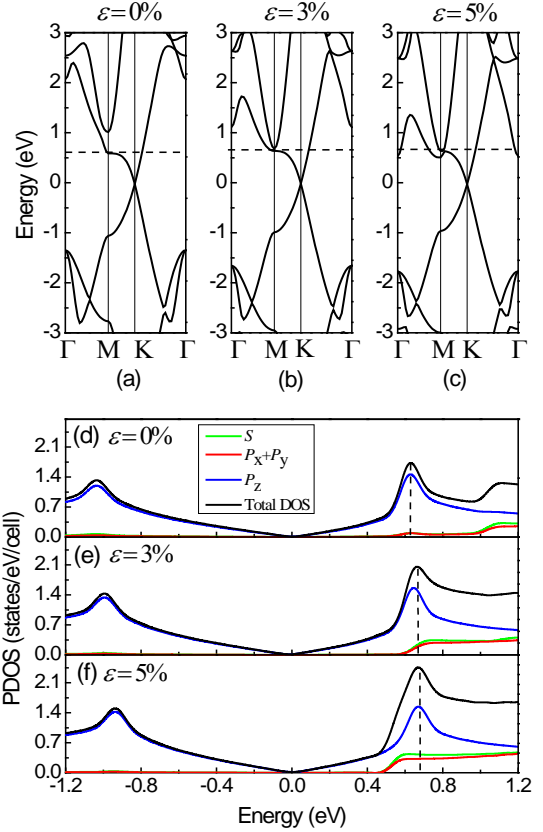

Fig. 1: (Colour on-line) Band structure of silicene near DP under the tension of (a) $\varepsilon = 0\%$, (b) $\varepsilon = 3\%$ and (c) $\varepsilon = 5\%$. Fermi level is set to zero. Projected Density of states for silicene under the tension of (d) $\varepsilon = 0\%$, (e) $\varepsilon = 3\%$ and (f) $\varepsilon = 5\%$. The dash line in band structure and PDOS is the location of Van Hove singularity whose corresponding $n_e$ is about $1.8\times10^{14}$ cm$^{-2}$, $2.0\times10^{14}$ cm$^{-2}$, $3.1\times10^{14}$ cm$^{-2}$ for tension of $\varepsilon = 0\%$, 3% and 5%, respectively.

Phonon dispersive relations of silicene under the tension of $\varepsilon = 0\%$, 3% and 5% are displayed in fig. 2. Optical phonons are noticeably softened as a result of weakening of Si-Si bonds due to tensile strain. After 5% tension is applied to unstrained silicene, the frequency of highest optical phonon at $\Gamma$ point deceases from 568 cm$^{-1}$ to 493 cm$^{-1}$. But the frequencies of lowest three acoustic modes along $M$-$K$ path slightly increase. We further expand the lattice constant $a_0$ by 15 % and confirm that the lattice of silicene is still stable. This structural flexibility makes tension to be preferred in adjusting the lattice constant of silicene while keeping its lattice structure stable.

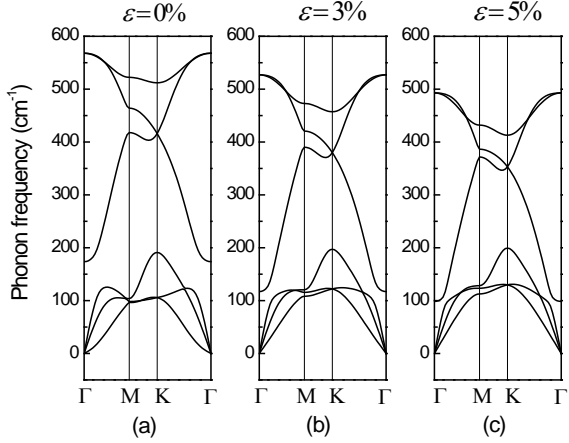

Fig. 2: Phonon dispersive relation of silicene under the tension of (a) $\varepsilon = 0\%$, (b) $\varepsilon = 3\%$ and (c) $\varepsilon = 5\%$, respectively.

Either BN substrate or hydrogen-processed Si surface has been proposed as a good candidate to preserve Dirac electrons in silicene [32]. One may be able to effectively reduce substrate's influence to band structure of silicene by choosing proper substrates in future synthesization. Within rigid band approximation [22], we tune the carrier concentration by shifting the Fermi energy $\epsilon_F$ away from DP while keeping the band structure and phonon dispersive relation of silicene unchanged. Considering that in a graphene based field-effect transistor, a doping level up to $4\times10^{14}$ cm$^{-2}$ for both electrons and holes has been reached by electrical gating [33], we calculated $\lambda$ and corresponding $T_c$ of silicene with the carrier concentration less than $4\times10^{14}$ cm$^{-2}$ under different tensions, The detail values of physical quantity in fig. (3) is available in the supplementary materials. The increase of carrier concentration can enhance $\lambda$ as shown in fig. 3(a). For the unstrained silicene, $\lambda$ is only 0.12 at the low electron doping of $n_e = 5\times10^{13}$ cm$^{-2}$. But at the high doping of $n_e = 3.5\times10^{14}$ cm$^{-2}$, $\lambda$ can reach 0.44. There exists a critical carrier concentration to produce a nonzero $T_c$ as shown in fig. 3(b). Both $\lambda$ and $T_c$ are larger in electron doping than that in hole doping at the same magnitude of carrier concentration. In the following, we will only discuss the electron doping.

The pronounced feature shown in fig. 3(b) is the clear promotion the $T_c$ of silicene by tensile strain. This effect is more notable at higher $n_e$. In the case of unstrained silicene, $T_c$ increases only 1.67 K even at the high doping of $n_e = 3.5\times10^{14}$ cm$^{-2}$. However, at 5% tension, $T_c$ can be pushed up to 16.4 K at the same $n_e$. The tension enhanced e-ph coupling can be related to the increase of DOS at the Fermi level. As the largest applied tension (5%) is experimentally accessible, the notable tension enhancement of SC of silicene may well be observed in future experiments.

To show the relative contribution of phonons with different frequencies to $\lambda$, we calculated the Eliashberg spectral function $\alpha^2 F(\omega)$ defined as [28]:

$$\alpha^2 F(\omega) = \frac{1}{2N_q} \sum_{q,\nu} \lambda_{q,\nu} \omega_{q,\nu} \delta(\omega_{q,\nu} - \omega). \quad (3)$$

We display the $\alpha^2 F(\omega)$ of silicene at $n_e = 1.5\times10^{14}$ cm$^{-2}$ and $3.5\times10^{14}$ cm$^{-2}$ under 5% tension in fig. 4(a) for clarity. The distribution of $\alpha^2 F(\omega)$ concentrates mainly at a high frequency zone around 470 cm$^{-1}$ and a low frequency zone around 130 cm$^{-1}$. As its lower frequency, the phonons at the low frequency zone make a more significant contribution to $\lambda$. One can clearly seen that increase of $n_e$ would especially increase the $\alpha^2 F(\omega)$ at the low frequency zone. Similar phenomenon was also found in silicene under other tensions.

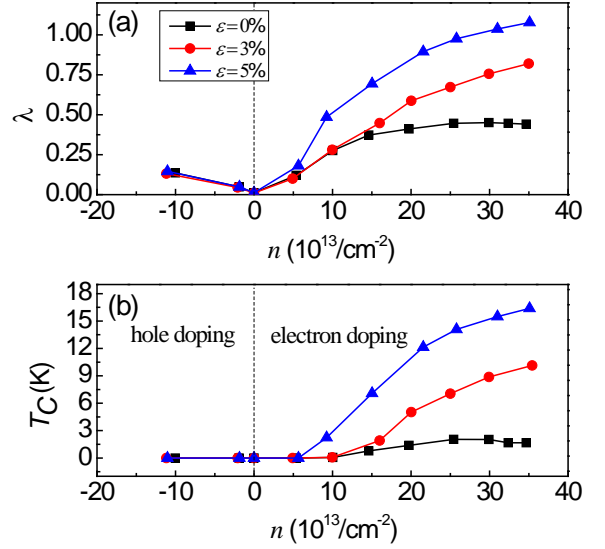

Fig. 3: (Colour on-line) Carrier concentration dependence of (a) $\lambda$ and (b) $T_c$ of silicene under different tensions. The black, red and blue line refers to silicene under the tension of $\varepsilon = 0\%$, 3 % and 5%, respectively. The positive value of $n$ represents electron doping ($n_e$) and the negative one represents hole doping ($n_o$).

In fig. 4(b), we show the $\alpha^2 F(\omega)$ of silicene at $n_e = 3.5\times10^{14}$ cm$^{-2}$ under different tensile strains, i.e. $\varepsilon = 0\%$, 3% and 5%. Same as the softening of high frequency phonons due to tension, which has been shown in fig. 2, the peak of $\alpha^2 F(\omega)$ at high frequency zone is shifted to smaller frequency. The $\alpha^2 F(\omega)$ at low frequency zone around 130 cm$^{-1}$ is clearly enlarged by tension. Tension enhancement of $\alpha^2 F(\omega)$ at low frequency zone was also found in silicene with other electron concentration $n_e$.

To reveal more clearly the relative contribution of different phonons to $\lambda$, we defined a integral function of $\lambda'(\omega)$ as:

$$\lambda'(\omega) = 2\int_0^\omega \frac{\alpha^2 F(\omega')}{\omega'} d\omega'. \quad (4)$$

The total EPC $\lambda$ is equal to $\lambda'(\omega \to \omega_{max})$, where $\omega_{max}$ is the highest frequency of phonon spectrum. At $n_e =$

$3.5\times10^{14}$ cm$^{-2}$, the $\lambda'(\omega)$ of silicene under the tension of $\varepsilon$ = 0%, 3% and 5% are displayed at fig. 4(c) for clarity. In the case of unstrained silicene, phonons at both low frequency zone and high frequency zone make a significant contribution to $\lambda$. But phonons at low frequency zone gradually dominate the $\lambda$ as the increase of tension.

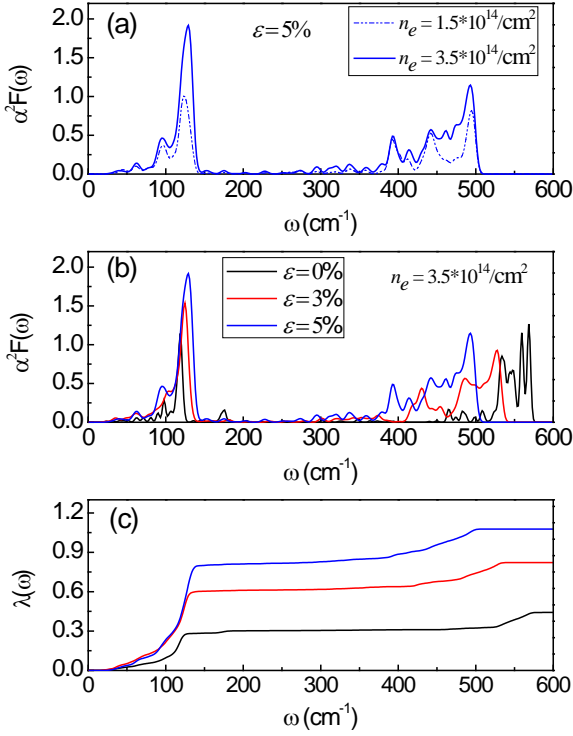

Fig. 4: (Colour on-line) (a) Eliashberg functions $\alpha^2F(\omega)$ of silicene at $n_e = 1.5\times10^{14}$ cm$^{-2}$ and $n_e = 3.5\times10^{14}$ cm$^{-2}$ under the tension of $\varepsilon$ = 5%. The dash and real blue line refers to silicene at $n_e = 1.5\times10^{14}$ cm$^{-2}$ and $n_e = 3.5\times10^{14}$ cm$^{-2}$, respectively. (b) The $\alpha^2F(\omega)$ of silicene at $n_e = 3.5\times10^{14}$ cm$^{-2}$ for the tension of $\varepsilon$ = 0%, 3% and 5%. (c) Function $\lambda'(\omega)$ for the tension of $\varepsilon$ = 0 %, 3% and 5% at $n_e = 3.5\times10^{14}$ cm$^{-2}$. Black, red and blue real line refers to silicene under the tension of $\varepsilon$ = 0%, 3% and 5%, respectively.

Further vibrational mode analysis shows that the peak of $\alpha^2F(\omega)$ at high frequency zone and low frequency zone are contributions mainly from the Si in-plane optical modes and the out-of-plane acoustic modes, respectively. The e-ph matrix element has a form of $\langle\psi_i|\delta V|\psi_j\rangle$, where both $\psi_i$ and $\psi_j$ are electronic states at the Fermi surface and $\delta V$ is the derivative of the self-consistent potential $V$ due to the phonon distortion. Polarization of out-of-plane acoustic phonons is perpendicular to the silicene sheet. It has been proved that the electronic states around DP in graphene are from $p_z$ orbitals of carbon atom, which is antisymmetric with respect to the plane of graphene. So the e-ph matrix element related to out-of-plane phonon modes is negligible in graphene for the symmetry constraint [34]. Silicene is not an ideal planar lattice for its low-buckled structure. The PDOS [see figs. 1(e) - 1(f)] also shows that there exists components arises from $s$, $p_x$ and $p_y$ orbitals of silicon atom in DOS induced by tension. The symmetry constraint existing in graphene is removed, and the out-of-plane phonons in silicene can contribute to the electron-phonon coupling.

To clearly reveal the origin of strong e-ph interaction in electron-doped silicene, we have calculated the nesting factor $X(\mathbf{q})$ which is defined as $X(\mathbf{q}) = \sum_{k,n,m}\delta(\epsilon_{k+q,m}-\epsilon_F)\delta(\epsilon_{k,n}-\epsilon_F)$. Larger $X(\mathbf{q})$ presents that more electronic states at the Fermi surface are connected by a wavevector $\mathbf{q}$. The $X(\mathbf{q})$ of silicene at doping of $n_e = 1.5\times10^{14}$ cm$^{-2}$ under the tension of $\varepsilon$ = 0% and $\varepsilon$ = 5% are shown in fig. 5(a) and 5(b), respectively. At the low doping level, electronic states at Fermi surface interact mainly with phonons at the center of BZ.

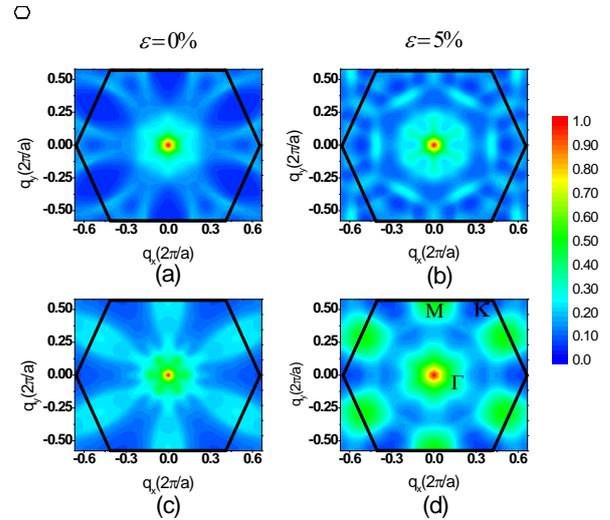

Fig. 5: (Colour on-line) Nesting factor $X(\mathbf{q})$ of silicene under the tension of (a) $\varepsilon$ = 0% and (b) $\varepsilon$ = 5% at $n_e = 1.5\times10^{14}$ cm$^{-2}$. Nesting factor $X(\mathbf{q})$ for the tension of (c) $\varepsilon$ = 0% and (d) $\varepsilon$ = 5% at $n_e = 3.5\times10^{14}$ cm$^{-2}$. The data have been normalized to the largest value.

The $X(\mathbf{q})$ of silicene at high doping of $n_e = 3.5\times10^{14}$ cm$^{-2}$ under the tension of $\varepsilon$ = 0% and $\varepsilon$ = 5% are shown in fig. 5(c) and 5(d), respectively. Obviously the $X(\mathbf{q})$ around $M$ point are greatly enhanced due to tension. By further comparing the $\lambda_{q,\nu}$ of each phonon for silicene under 5% tension, it is found that the out-of-plane phonon modes around $M$ point have the largest $\lambda_{q,\nu}$, which is consistent with the results of $X(\mathbf{q})$. Under 5% tension [see fig. 1(c)], it is clearly seen that the bands at $\Gamma$ point and $M$ point are in the same energy range. Tension opens new e-ph scattering channels between electronic states in these two points. Considering that the bands above DP decrease continuously with tension larger than 5%, $T_c$ will increase further.

In conclusion, the biaxial strain effects on the electron-phonon coupling in electron-doped silicene have been extensively studied based on first-principles methods within rigid band approximation. Tensile strain can not only increase density of states at Fermi level but also create new channels for e-ph coupling. That leads to a strong e-ph interaction occurring between the out-of-plane phonon modes and electronic states at the Fermi surface. The $T_c$ of electron-doped silicene can be increased to the value exceeding 10 K by 5% tensile strain.

∗∗∗

We are grateful to Prof. Feng liu for fruitful discussions. This work were supported by MOST Project of China (Grants No. 2014CB920903, No. 2011CBA00100), Natural Science Foundation of China (Grants No. 11225418 and 11174337), and Specialized Research Fund for the Doctoral Program of Higher Education of China (Grants No. 20121101110046).